\documentclass{iau}
\usepackage{graphicx}
\def\apj{{ApJ}}                 
\def\apjl{{ApJ}}                
\def\mnras{{MNRAS}}             
\def\prd{{Phys.~Rev.~D}}        

\title[Testing Gravity using Void Profiles]
  {Testing Gravity using Void Profiles}
\author[Cai et al.]
 {Yan-Chuan~Cai$^1$, Nelson Padilla$^{2,3}$, Baojiu~Li$^1$}
\affiliation{$^1$Institute for Computational Cosmology, Department of Physics, University of Durham, South Road, Durham DH1 3LE, UK \\ email: {\tt y.c.cai@durham.ac.uk} \\[\affilskip]
$^2$Instituto de Astrof\'isica, Pontificia Universidad Cat\'olica, Av. Vicu\~na Mackenna 4860, Santiago, Chile. \\
$^3$Centro de Astro-Ingenier\'ia, Pontificia Universidad Cat\'olica, Av. Vicu\~na Mackenna 4860, Santiago, Chile.
}
\pubyear{2014}
\volume{308}  
\pagerange{XXX--XXX}
\jname{The Zeldovich Universe}
\editors{Rien van de Weygaert, Sergei Shandarin, \\
                     Enn Saar \& Jaan Einasto, eds.}

\begin{document}

\maketitle

\begin{abstract}
We investigate void properties in $f(R)$ models using N-body simulations, focusing on their differences from General Relativity (GR) and their detectability. 
In the Hu-Sawicki $f(R)$ modified gravity (MG) models, the halo number density profiles of voids are 
not distinguishable from GR. In contrast, the same $f(R)$ voids are more empty of dark matter, and their profiles are steeper. 
This can in principle be observed by weak gravitational lensing of voids, for which the combination of a spectroscopic redshift and a lensing photometric redshift 
survey over the same sky is required. Neglecting the lensing shape noise, the $f(R)$ model parameter amplitudes $|f_{R0}|=10^{-5}$ and $10^{-4}$ may 
be distinguished from GR using the lensing tangential shear signal around voids by 4 and 8$\sigma$ for a volume of  1~(Gpc/$h$)$^3$. The line-of-sight projection 
of large-scale structure is the main systematics that limits the significance of this signal for the near future wide angle and deep lensing surveys. For this reason, 
it is challenging to distinguish $|f_{R0}|=10^{-6}$ from GR. We expect that this can be overcome with larger volume. The halo void abundance being smaller and the 
steepening of dark matter void profiles in $f(R)$ models are unique features that can be combined to break the degeneracy between $|f_{R0}|$ and $\sigma_8$. 
\end{abstract}

\firstsection 
\section{Introduction}
Scalar-field models of modified gravity can mimic the late-time accelerated expansion of the Universe without invoking a
cosmological constant, but usually, extra long-range fifth forces are introduces. These models can still in principle pass local 
tests of gravity via certain screeening mechanisms. The Vainshtein \cite[(Vainshtein 1972)]{Vainshtein1972} or chameleon 
\cite[(Khoury \& Weltman, 2004)]{kw2004} mechanisms are two of the typical ones via which the fifth forces are suppressed in high density 
regions like dark matter haloes and the local Solar system. To distinguish these models from GR, 
it is therefore important to investigate the under dense regions, where the differences between MG and GR may be larger. 
A study using spherical evolution model by  \cite[(Clampitt et al. 2013)]{Clampitt2013} has shown that in chameleon 
models, fifth forces in voids are repulsive. Voids are driven by the fifth force to expand faster and grow larger. 
In this work, we use N-body simulations to investigate observables for this phenomena, using the 
Hu-Sawicki $f(R)$ models \cite[Hu \& Sawicki (2007)]{HuSawicki2007} as an example. We will focus on void profiles in this proceeding. 
A detailed analysis including void abundances is presented in \cite[(Cai et al. 2014)]{CPL}.

\section{Simulations and void finding}
Simulations of the  Hu-Sawicki $f(R)$ models with model parameter amplitudes of 
$|f_{R0}|=10^{-6}$ (F6), $10^{-5}$(F5) and $10^{-4}$(F4) are employed, 
where the model F6 has the weakest coupling strength between the scalar field and the density 
and is the most similar to GR. These simulations are 
performed using the {\sc ecosmog} code \cite[(Li et~al. 2012)]{lztk2012} with the same initial 
conditions and the same background expansion history as that of a GR $\Lambda$CDM model, 
which make it straightforward for comparison. Cosmological parameters of the simulations are: 
$\Omega_m = 0.24$, $\Omega_\Lambda = 0.76$, $h = 0.73$, $n_s = 0.958$ and $\sigma_8 = 0.80$, 
where $\Omega_m$ is the matter content of the universe, $\Omega_\Lambda $ is the effective dark energy density, 
$h$ is the dimensionless Hubble constant at present, $n_s$ is the spectral index of the primordial power spectrum 
and $\sigma_8$ is the linear root-mean-squared density fluctuation in spheres of radius 8${\rm Mpc}/h$. 
The simulations have the boxsize of 1Gpc/$h$ and $1024^3$ particles. 

Dark matter haloes are found using the spherical 
overdensity code AHF\cite[(Knollmann \&  Knebe 2009)]{arf}. We use haloes above the minimal 
mass $M_{\rm min} \sim 10^{12.8}M_{\odot}/h$ so that each halo has at least 100 particles. For fair comparison, 
we adjust this value slightly for different models such that the number density of haloes are the same. 
This makes sure that the void population found for different models are least affected by possible 
differences of shot noise. In this sense, any model differences we find for the 
void properties are perhaps conservative. Voids are defined using an improved version of the spherical 
overdensity algorithm of \cite[Padilla  et~al. (2005)]{Padilla2005}.

We define voids in halo fields as they are related to galaxy clusters and groups, which are observational 
meaningful. We call them halo voids. Using tracers of dark matter to define voids, however, will suffer from the 
effect of sparse sampling. To overcome this, we introduce the variance of the distances for the nearest four halos from each 
void center, $\sigma_4$ as a free parameter to control the quality of halo voids. Voids with relatively small value 
(0.2) of $\sigma_4$ are chosen such that they are close to spherical in shape. Sub-voids that are 
100\% overlapped with a main void are excluded by default. Details of void finding can be found in 
\cite[(Padilla  et~al. 2005)]{Padilla2005} and \cite[(Cai et al. 2014)]{CPL}.

\begin{figure}
\advance\leftskip -0.0cm
\scalebox{0.345}{
\includegraphics[angle=0]{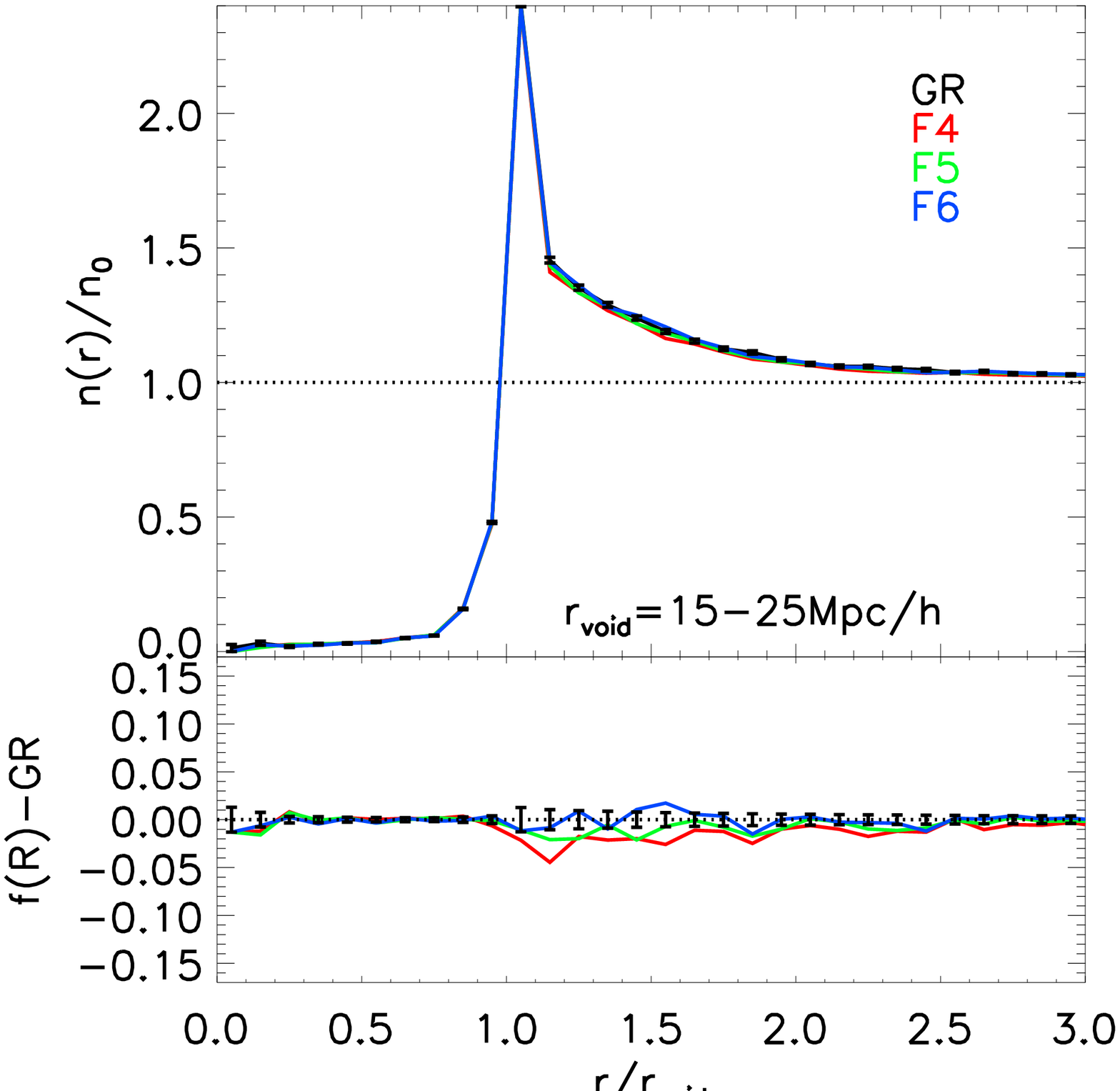}
\includegraphics[angle=0]{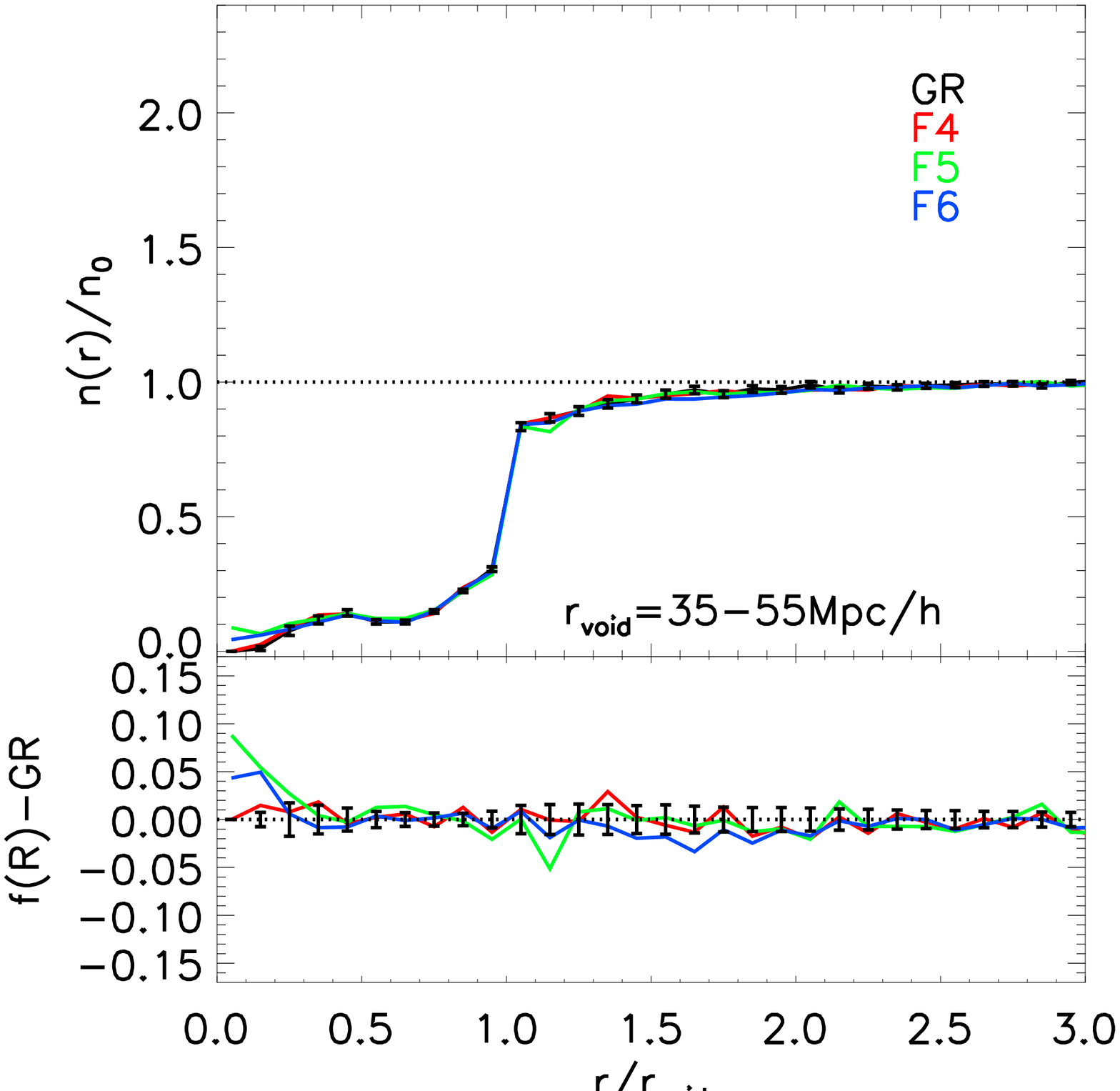}}
\caption{Top panels: the halo number density profiles of voids using haloes above the minimum halo mass of $M_{\rm min}\sim 10^{12.8} M_{\odot}/h$ from 
simulations of different models as labelled in the legend. $M_{\rm min}$ is slightly different from $10^{12.8} M_{\odot}/h$ in the $f(R)$ models so that the number of haloes for different models are the same. Error bars shown on the black line (GR) are the scatter about the mean for voids at $15~{\rm Mpc}/h<r_{\rm void}<25~{\rm Mpc}/h$ 
(left) and at $35~{\rm Mpc}/h<r_{\rm void}<55~{\rm Mpc}/h$ (right) found within the 1(Gpc/$h$)$^3$ volume. There are [6038, 5946, 6096, 6307] (left)  
and [296, 323, 319, 261] (right) voids in GR, F6, F5 and F4 models passing the selection criteria. Bottom panels: differences of halo number density profiles of 
voids between $f(R)$ models and GR.}
\label{fig:Haloprofile}
\end{figure}

\begin{figure}
\begin{center}
\advance\leftskip -0.0cm
\scalebox{0.345}{
\includegraphics[angle=0]{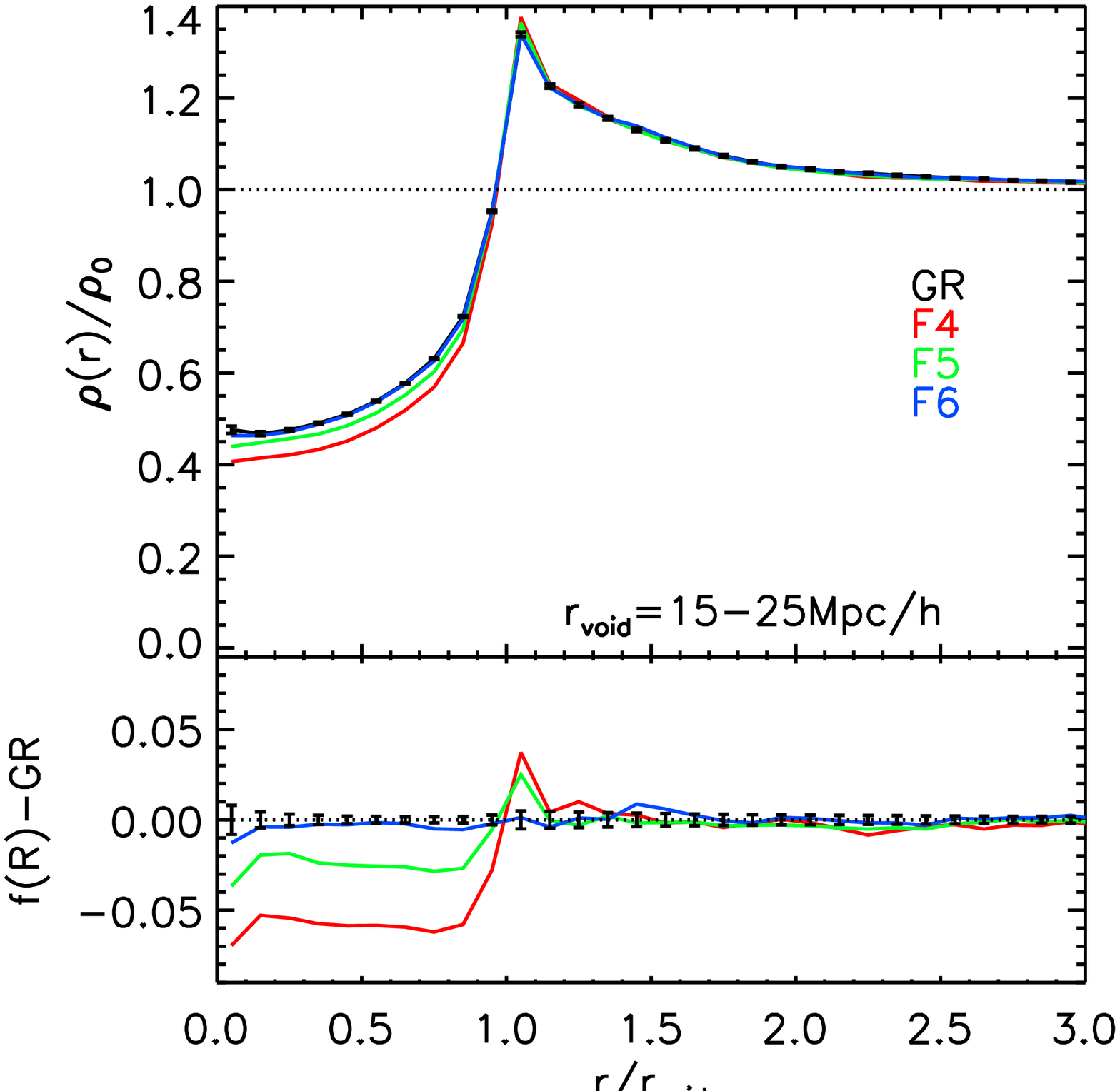}
\includegraphics[angle=0]{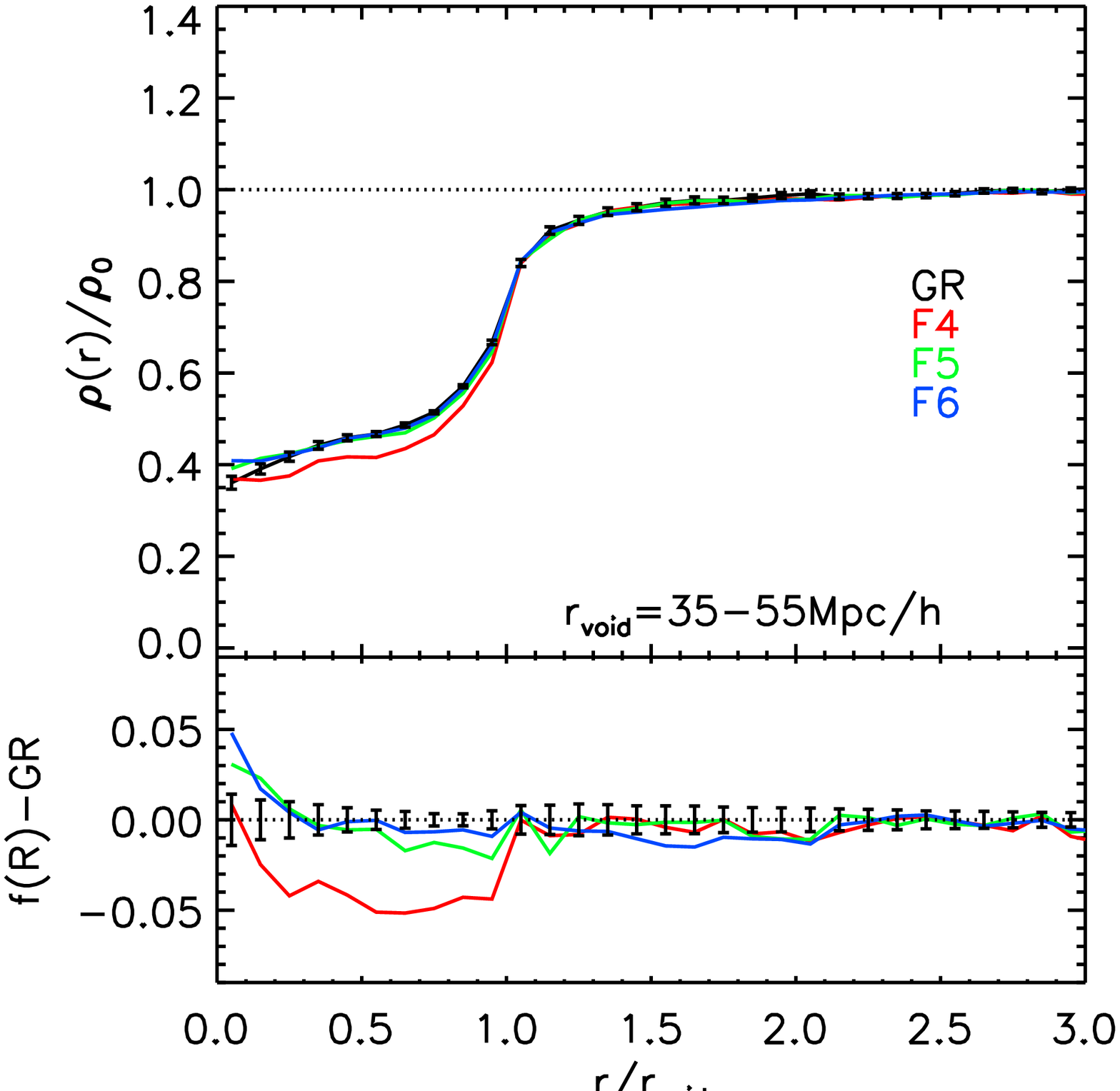}}
\caption{Similar to Fig.~\ref{fig:Haloprofile} but showing void density profiles measured using all dark matter particles from simulations of 
different models as labelled in the legend. Voids are defined using halo number density fields, which are the same as those being used to make 
Fig.~\ref{fig:Haloprofile}.}
\label{fig:Dprofile}
\end{center}
\end{figure}
\section{Void profiles}
With void centers found in both $f(R)$ and GR simulations, we measure their halo number density profiles as 
well as the dark matter density profiles. Results are presented in 
Fig.~\ref{fig:Haloprofile} and Fig.~\ref{fig:Dprofile}. The left and right panels are void profiles with small radius and large radius, which qualitatively 
correspond to two different types of voids, void-in-cloud (voids in overdense environment) and void-in-void 
(voids in underdense environment) \cite[(Sheth \& van de Weygaert, 2004)]{Sheth2004}. Profiles on the left have overdense ridges at 1$\times$ void radius $r_{\rm void}$, but not for 
those on the right. These are qualitatively as expected from the dynamics of void evolution. Small voids are likely to have been through more shell crossings at their walls than large ones. 
They are also consistent with the expansion velocities of voids. Void-in-cloud has a regime of infall at about $r>1.5\times r_{\rm void}$ but the outflow of 
mass persist at all scales for void-in-void up to 3 times of void radius [see \cite[(Cai et al. 2014)]{CPL} for more details]. In general, the halo number density profiles are 
steeper than the dark matter ones at $\sim r_{\rm void}$.

Most interestingly, we find that the halo number density profiles of voids are not distinguishable for different models (top panels of Fig.~\ref{fig:Haloprofile}). In contrast, 
the dark matter density profiles are deeper in $f(R)$ within $r>1\times r_{\rm void}$ (Fig.~\ref{fig:Dprofile}). Also, the dark matter overdense ridges at $r \sim r_{\rm void}$, if any, 
are sharper in $f(R)$ models. 

It is somewhat counter-intuitive but perhaps not surprising that little model difference is shown in 
halo number density profiles, because the fifth force is suppressed in overdense regions in $f(R)$. The spatial distribution of massive 
haloes are perhaps not so different between $f(R)$ and GR. The dark matter void profiles being steeper in $f(R)$ is consistent with 
the analytical results of \cite[(Clampitt et al. 2013)]{Clampitt2013} that the repulsive fifth force drives voids in $f(R)$ to grow larger and emptier. 
It is also reasonable that once the growth of voids are restricted by their environments, mass start to accumulate at the edges of voids, and 
the overdense ridges will grow sharper. This effect is stronger in $f(R)$ thanks to the repulsive fifth force in voids.

\begin{figure}
\begin{center}
\scalebox{0.345}{
\includegraphics[angle=0]{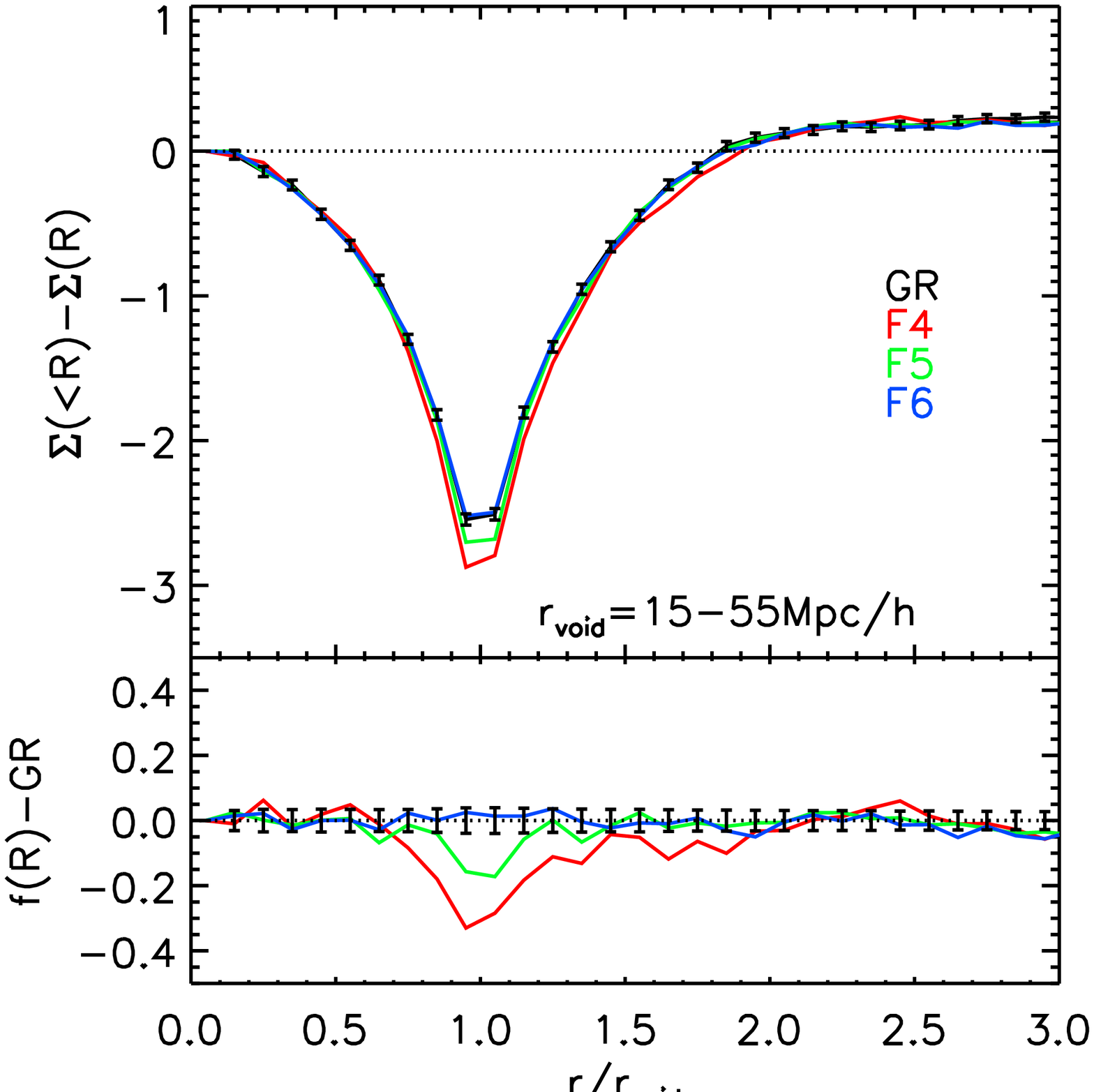}
\includegraphics[angle=0]{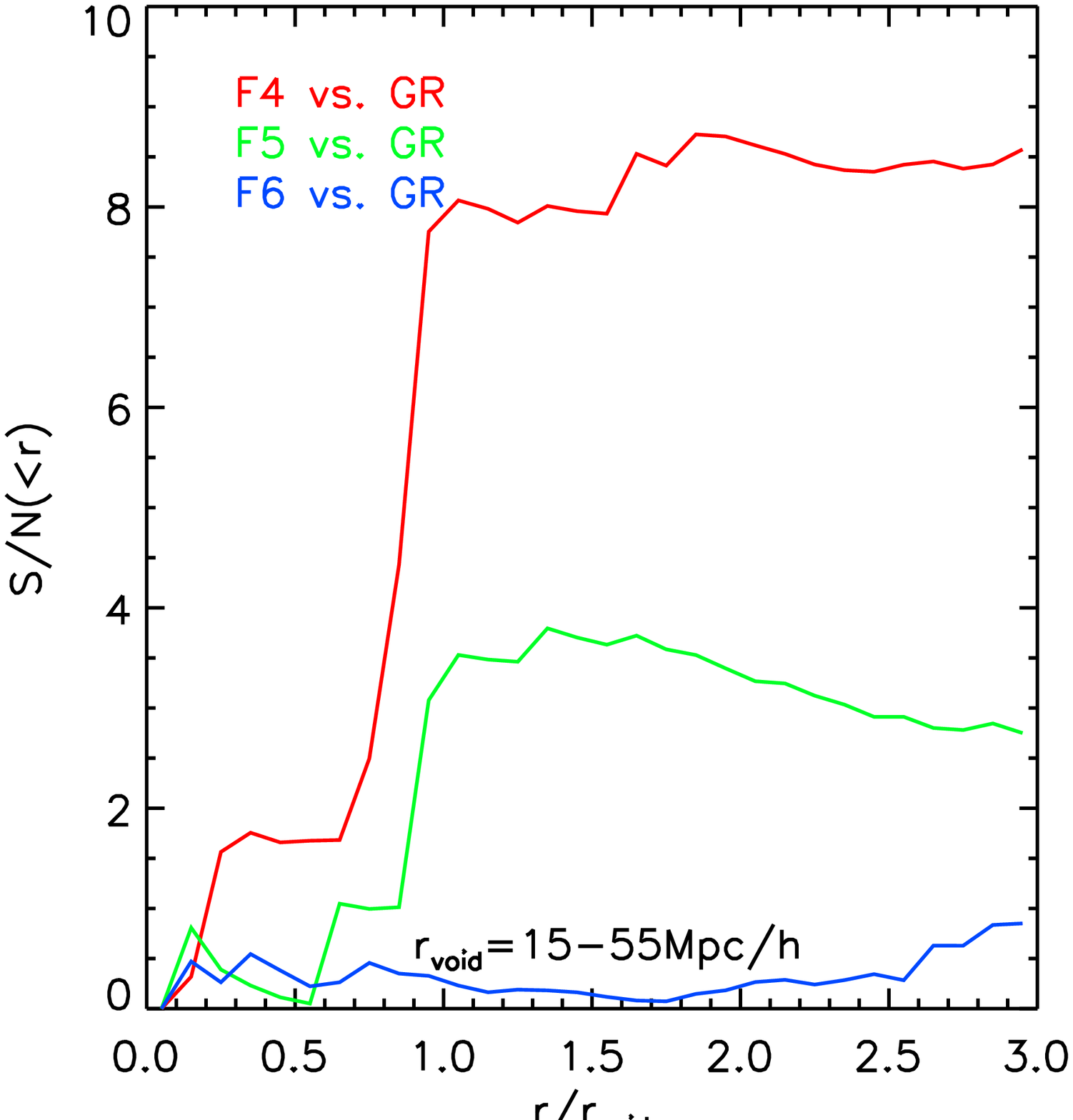}}
\caption{
Left: like Fig. \ref{fig:Dprofile} but showing the lensing tangential shear profiles from stacking all voids with $15<r<55$ Mpc/$h$. 
They are projected over two times the void radius along the line of sight. 
$\Sigma(<R)-\Sigma(R)$ is proportional to the surface mass density within the projected radius of $R$ to which we subtract the surface mass density at $R$. 
Right: the corresponding cumulative (from small to large radius) S/N for the differences between GR and $f(R)$ models.}
\label{fig:Shearprofile}
\end{center}
\end{figure}
\subsection{Gravitational lensing of voids}
The differences between $f(R)$ and GR in dark matter void profiles can in principle be observed using 
gravitational lensing of voids. This requires the overlap of a spectroscopic redshift survey (spec-$z$) and 
a (deeper) lensing photometric redshift survey (photo-$z$) over the same sky. Voids will be identified 
in the spec-$z$ survey while their gravitational lensing effect on the background galaxy images can 
be detected in the photo-$z$ survey via stacking of void centres. The lensing tangential shear profiles of 
the background galaxies are associated with the excess of the projected mass density along the line of sight, 
\begin{eqnarray}
\Delta \Sigma(R)=\Sigma(<R)-\Sigma(R),
\end{eqnarray}
where $\Sigma(R)$ and $\Sigma(<R)$ are the projected surface densities 
around the centre of a void at the projected distance of $R$ and within $R$. 
It can be used to measure the void density profile \cite[(Krause et~al. 2013]{Krause2013}; \cite[Higuchi et al. 2013]{Higuchi2013}; 
\cite[Clampitt \& Jain 2014)]{Clampitt2014}. This is sensitive to the slope of the mass density. 
Therefore, the peak of the lensing signal for voids is at 
$\sim r_{\rm void}$ where the slope of the matter density is the largest, shown in Fig.~\ref{fig:Shearprofile}. Neglecting the lensing shape noise, 
F5 and F4 can be told apart from GR by 4 and 8$\sigma$ for the 1(Gpc/$h$)$^{3}$ volume. 
F6 is not distinguishable from GR due to the line-of-sight projection of large-scale structure. 

We find sub-voids are useful to help increasing the lensing S/N, though the S/N per void is not as 
great as that of the main voids. With all sub-voids included in our volume, the number of voids increases by 
approximately $76\%$. The S/N for F5 and F4 are increased to 7 and 12, but there is no increase of S/N  for F6. 
The S/Ns are degraded if we integrate the projected mass density to larger line-of-sight distances. For example, 
increasing the line-of-sight projection from 2 to 6 times of void radius decrease the S/Ns by about 30\%. 

The above forecast may be somewhat optimistic as the lensing shape noise and other systematics are neglected.
However, at the relatively large radius ($\sim r_{\rm void}$), which are the most interesting to distinguish $f(R)$ models from GR,  the lensing shape noise is expected to be 
sub-dominant for DEFT Stage IV type of deep imaging survey\cite[(Albrecht et~al. 2006)]{Albrecht2006}, see \cite[(Krause et~al. 2013]{Krause2013}; \cite[Higuchi et al. 2013)]{Higuchi2013} 
for quantitative examples. The above forecast is based on a volume of 1(Gpc/$h$)$^{3}$. The current BOSS DR11 CMASS sample has an effective volume of 6.0 (Gpc/$h$)$^3$ \cite[(Anderson et al. 2011]{BOSS1}; \cite[Beutler et al. 2013]{Beutler2013}; 
\cite[S{\'a}nchez et al. 2014)]{Sanchez2014}.
In principle, the significance level should increase by a factor of 2.4 if the BOSS DR11 CMASS sample is used, on condition that deep lensing image data on the same sky is available. 
The future EUCLID survey \cite[(Laureijs et al. 2011)]{EUCLID} is expected to have an effective volume of $\sim$20 (Gpc/$h$)$^3$, a factor of 4.4 improvement is expected in this case.

\section{Conclusions and discussion}
Using simulations of $f(R)$ models, we have found that voids are emptier than that in GR 
in terms of dark matter. However, the void profiles of tracers (haloes) are not necessary distinguishable between $f(R)$ and GR. 
Moreover, the halo number density profiles of voids are very different from that of dark matter, the former being sharper. 
This is true even in GR. It rings an alarm that voids found using tracers are not necessary the same as that of dark matter. Note that other authors using 
different void finding algorithms may conclude differently, see for example \cite[(Sutter et al. 2014)]{SutterLavaux2014}. 

We have found that two types of voids, void-in-cloud and void-in-void, are separable using their radii, the latter tend to be larger. 
Their profiles are different in that the former have developed over dense ridges but not for the latter. From our prospective, 
it is perhaps unlikely that the void profile takes the same form, and can simply be rescaled only by the void radius for these two different types. 
For void-in-cloud, an additional parameter is needed to describe the height of the overdense ridge. This seems different from the results of 
\cite[(Nadathur et al. 2014)]{Nadathur2014}, but again, they are using ZOBOV \cite[(Neyrinck et al. 2005)]{Neyrinck05} to find voids. Also, the voids they found are from (mock) LRG galaxies and their void sizes 
are relatively large compared to ours.
 
Using halo voids to study their dark matter density profiles has the observational implications in that voids are 
usually found using tracers. The steepening of the underlying dark matter void profile in $f(R)$ models over that of GR 
induces stronger lensing tangential shear signals at about $1\times r_{\rm void}$. 
Measuring the model differences associated with these voids requires the combination of a spec-$z$ survey 
and a photo-$z$ survey. This adds value to the idea of combining surveys on top of systematic calibration and canceling of cosmic variance 
\cite[(Zhang et al. 2007]{Zhang2007}; \cite[McDonald \& Seljak, 2009]{McDonald2009}; \cite[Bernstein \& Cai, 2011]{BC2011};  \cite[Cai \& Bernstein, 2011]{Cai2012}; 
\cite[Gazta{\~n}aga et al. 2012]{Gaztanaga2012}; \cite[de Putter et al. 2013]{dePutter2013}; \cite[Kirk et al. 2013)]{Kirk2013}.
Neglecting lensing shape noise, which is expected to be sub-dominant 
for near future deep imaging surveys, F5 and F4 can be told apart from GR by 4 and 8$\sigma$. Line-of-sight projections of large-scale 
structure set limit on the constraining power. For this reason, it is challenging to distinguish F6 using the (1Gpc/$h$)$^3$ volume.

We caution that the steepening of void profiles may also be expected in the same $\Lambda$CDM model with a higher $\sigma_8$. 
In this sense, there may be a degeneracy between the $|f_{R0}|$ parameter with $\sigma_8$. This may be possible to be broken 
using the void abundance measurement. In \cite[(Cai et al. 2014)]{Cai2014}, the halo void abundances 
are found to be smaller for large voids in $f(R)$ models compared to that in GR. This counters the trend in $\Lambda$CDM with a higher $\sigma_8$. 
Therefore, the combination of void abundances and profiles may be a powerful tool for constraining gravity.

\end{document}